# Reality of the wave function
# and quantum entanglement

Mani Bhaumik[1]
Department of Physics and Astronomy,
University of California, Los Angeles, USA.90095

**Abstract**

The intrinsic fluctuations of the underlying, immutable quantum fields that fill all space and time can support the element of reality of a wave function in quantum mechanics. The mysterious non-locality of quantum entanglement may also be understood in terms of these inherent quantum fluctuations, ever-present at the most fundamental level of the universe.

## 1. Introduction

The reality of the wave function in quantum mechanics has been controversial ever since Schrödinger introduced it in his equation of motion. Nevertheless, it would be reasonable to embrace at least a tentative conception of the quantum reality since quantum formalism invariably underpins the reality of our entire objective universe.

[1] e-mail: bhaumik@physics.ucla.edu

In a recent article [1], it was pointed out that the wave function plausibly represents a structural reality of the very foundation of our universe. In brief, contemporary *experimental observations* supported by the quantum field theory demonstrate the universal presence and immutability of the abiding quantum fields at the ultimate level of reality as well as some definitive confirmation of the existence of inherent ceaseless fluctuations of these fields. Let us now explore how an elementary particle like electron can have its inescapable associated wave as a result of the incessant fluctuations of the primary quantum field.

According to QFT, an electron represents a propagating discrete quantum of the underlying electron field. In other words, an electron is a quantized wave (or a ripple) of the electron quantum field, which acts as a particle because of its well-defined energy, momentum, and mass, which are conserved fundamentals of the electron. However, even a single electron, in its reference frame, is never alone. It is unavoidably subjected to the perpetual fluctuations of the quantum fields.

To begin with, the quantum fluctuations continually and prodigiously create virtual electron-positron pairs in a volume surrounding the electron. Even though each pair has a fleeting existence, of the order of $10^{-21}$ second, on an *average* there is a very significant amount of these pairs to impart a remarkably sizable screening of the bare charge of the electron.

Other quantum fluctuations cause the electron to spontaneously emit a virtual photon, which is a disturbance in the electromagnetic field due to the presence of the electron. To conserve momentum, the electron would recoil with momentum equal and opposite to that of the photon. A quantum fluctuation of energy ΔE will provide the kinetic energy for the recoil of the electron as well as the energy of the photon for a time ΔT ∼ ℏ/ΔE. During this transitory moment, the electron by creating a disturbance in the photon field becomes a disturbed ripple itself and therefore ceases to be a normal particle. However, the combination of the two disturbances, or the virtual particles as they are commonly called, retains the characteristics of the real electron.

The disturbance in the photon or the electromagnetic field in turn can cause disturbances in other electrically charged quantum fields, like the muon and the various quark fields. Generally speaking, in this manner, every quantum particle spends some time as a mixture of other particles in all possible ways. However, the *combination of the disturbances* in the electron field together with those in all the other fields always maintains a well-defined energy and momentum with an electron mass, since they are conserved quantities for the electron as a particle.

 Although individual disturbances in the fields or the virtual particles due to quantum fluctuations have an ephemeral existence, on the average there ought to be a significant number of disturbances present at any particular time, just as in the case of the momentarily existing pairs providing the significant screening for the bare charge of the electron. This would make the momentum of the electron appear as a function of time, rendering its motion jittery in the very short time scales characteristic of the quantum fluctuations. The cumulative effect of the fluctuations on the particle should conform to a conceivable pattern of the quantum fluctuations, which could result in the wave or the wave packet that continually accompanies the particle.

The effect of the quantum fluctuations resulting in a wave packet associated with the particle can perhaps be better envisioned by an alternative approach. Recalling that an electron is a quantized ripple of the electron quantum field, it acts as a particle because it travels with its conserved quantities always sustained holistically as a unit. However, due to interactions of the particle with all the other possible quantum fields, the ripple in fact is very highly disturbed, which can be expressed by Fourier analysis into a weighted linear combination of simple wave forms like trigonometric and Gaussian functions. The result would be a wave packet or a wave function that represents a fundamental reality of the universe.

It can therefore be suggested that the random disturbances caused by the inherent quantum fluctuations of the underlying field is the reason for a quantum particle like an electron to be always associated with a wave and also this lends foundation to the element of reality of Bohm's quantum $\psi$–field [2]. In support of this notion, the holistic nature of the wave or the wave function is presented as evidence. In a measurement, this holistic nature becomes obvious since the appearance of the particle in one place prevents its appearance in any other place.

Contrary to the waves of classical physics, the wave function cannot be sub-divided during a measurement. This is because *only the combination* of all the disturbances comprising the wave function possesses a well-defined energy and momentum with the mass of the particle. Consequently, the wave function disappears everywhere else except where it is measured. *This experimental fact could provide a solution to the well-known measurement paradox.* It has been very difficult to understand why, after a unitary evolution, the wave function suddenly collapses upon measurement or a similar reductive interaction. The holistic nature of the wave function described above seems to offer a plausible explanation. Parts of the wave function that might spread to a considerably large distance can also terminate instantaneously by the process displayed in quantum entanglement discussed in section 5.

*Thus, the profound fundamentals of our universe appear to support the objective element of reality of the wave function, which represents a natural phenomenon and not just a mathematical construct.* We also observe that while the wave nature predominates as a very highly disturbed ripple of the quantum field before a measurement, the particle aspect becomes paramount upon measurement.

In this communication we present further substantiation of the premise that a wave function naturally arising due to quantum fluctuations can be used to represent a fundamental particle, extending it to include multi particle systems with a particular emphasis on quantum entanglement.

## 2. Hydrogen atom

Despite the roiling ocean of quantum fluctuations, some order can be found in the midst of all the unpredictability. A familiar example is the decay of radioactive atoms. The instance of decay for any particular atom is completely spontaneous and totally unpredictable. But for a sufficient number of these atoms, the time required for the decay of half of them is evidently calculable. Likewise, the random quantum fluctuations of the fields in any space time element can be embodied in a wave function. Although this can perhaps be accomplished in alternative ways, we will use the linear superposition of the quantum harmonic oscillator wave functions, which are Gaussian, as in reference [1].

Accordingly, with the necessary adjustment of coefficients in each case, the wave function $\psi$ of a particle as well as of the vacuum quantum fluctuations can be written as

$$\psi = \Sigma_i \ c_i \ \psi_i$$

Since all the $\psi_i$ are solutions of the Schrödinger equation, their linear superposition with the appropriate adjustments of all the coefficients $c_i$ should satisfy each of the constraints applicable for a wave function to be used in the Schrödinger equation. Specifically, $\psi$ and its first derivative would be finite and continuous everywhere. Additionally, $\psi$ would be normalizable so that the sum of the probabilities over all space would be one. Thus,

$$\int_{-\infty}^{+\infty} \psi * \psi = 1$$

Such a wave function $\psi$ embodying the quantum fluctuations can be utilized for the Schrödinger equation of a quantum system. For example, it can be used for the time independent Schrödinger equation of the extensively studied hydrogen atom:

$$-\frac{\hbar^2}{2m} \nabla^2 \psi + V\psi = E\psi$$

As usual, the energy levels of the hydrogen atom can be found by solving this equation with the well-established procedure, drawing on the principle of separation of variables and employing the applicable boundary conditions.

In the same way, other stationary states of quantum system involving a single particle can be calculated and their wave function construed in terms of the reality of the primary quantum fluctuations. In a previous article [1], it was shown that time dependent single particle quantum phenomena are explainable in this way as well. Such phenomena would include the marvel of quantum tunneling [2].

## 3. Two-particle System

We have described [1] how the essential features of a one particle quantum system can be interpreted by using the objectively real field. This field determines the magnitude of a quantum potential Q that provides the quantum force for the jittery temporal motion of the particle. Thus it is possible to understand the one particle quantum system without the necessity of any significant change in our overall understanding of space, time, and causality.

However, when we attempt to understand the many-particle system in this way, we confront a radically new concept. If we have, for example, many electrons, they are all quantized waves of the same underlying electron field, teeming inherently with frantic fluctuations. Under these circumstances, a quantum interconnectedness of all the particles introduces the notion of a *wholeness* of the entire quantum system. Of particular significance is the discovery of quantum entanglement of particles, a concept coined by Schrödinger, which ensues when the quantum state of each particle must be described relative to the other. Penrose finds this extremely puzzling, stating [3], "It is remarkable that we seem to have to turn to something so esoteric and hidden from view when, for many particle systems, almost the *entire* 'information' in the wave function is concerned with such matters!"

Observation of the distinctly nonlocal nature of quantum entanglement that has now gained wide acceptance became feasible only after John Bell masterfully formulated his famous inequality relation [4]. It has now become almost a routine to demonstrate that when some property of one of the particles in an entangled pair is measured, the other particle instantaneously responds irrespective of how far the two particles may be separated in space.

This has opened up the possible use of quantum entanglement in a variety of novel applications such as quantum cryptography, quantum computation, and quantum teleportation, which have become areas of very active research. Therefore a comprehensive and vivid understanding of the phenomenon would be very useful.

The incredibly perplexing effect of non-locality was famously called by Einstein "spooky action at a distance." Despite their relatively recent acceptance as a tenet, these properties of non-locality and entanglement are still baffling to physicists. We present here a plausible explanation in terms of the manifest fluctuations of the underlying, abiding quantum field.

We start with a two particle system. Generalization of the results to a many particle system will follow from this in a rather straightforward manner. Let us designate $\psi_1$ and $\psi_2$ as the *Bohm fields* for particle 1 and 2 respectively, resulting from the intrinsic fluctuations of the underlying quantum field. If there are no interactions between the particles as well as their $\psi$ fields, the combined wave function can be written as a product of their individual wave functions. However, as a consequence of quantum interconnectedness, the particles do interact and their wave function would be an appropriate superposition of the product states that are not separable. The wave function then consists of the coordinates of both particles. Accordingly, $\Psi(X_1, X_2, t)$ for two nonrelativistic particles of equal mass with no spin satisfies the time dependent Schrödinger equation,

$$i\hbar \frac{\partial \Psi}{\partial t} = -\frac{\hbar^2}{2m} (\nabla_1^2 + \nabla_2^2)\Psi + V\Psi$$

where $\nabla_1$ and $\nabla_2$ refer to particles 1 and 2 respectively, while $X_1$ and $X_2$ each represents three space coordinates.

Expressing $\Psi$ as $\Psi = R(X_1, X_2) \exp(iS(X_1, X_2)/\hbar)$ where $R$ and $S$ are both real with $R^2 = P = \Psi^*\Psi$, $v_1 = \nabla_1 S/m$ and $v_2 = \nabla_2 S/m$, and substituting it in the Schrödinger equation, we get two equations. One of them is the quantum mechanical equivalent of the Hamilton-Jacobi equation,

$$\frac{\partial S}{\partial t} + \frac{(\nabla_1 S)^2}{2m} + \frac{(\nabla_2 S)^2}{2m} + V + Q = 0$$

where the quantum potential $Q$ is

$$Q = -\frac{\hbar^2}{2m} \frac{(\nabla_1^2 + \nabla_2^2)R}{R}.$$

The wave function $\Psi$ now depends on the six variables $X_1$ and $X_2$ constituting the coordinates of the two particles and on time $t$. Obviously, $\Psi(X_1, X_2, t)$ can no longer be deemed as a field in typical three dimensional spaces. Instead, it is a function expressed in the configuration space of the two particles.

Since the quantum potential depends on $\Psi(X_1, X_2, t)$, it is therefore determined by the quantum state of the system as a whole. This suggests the quantum potential Q directs the quantum interaction between the particles in a reflexively interrelated way. Also, in the expression for Q, R appears both in the numerator and the denominator. Therefore, multiplying the wave function by a constant does not change Q, which thus does not fall off with distance. The significance then is that the two particles can remain coupled at arbitrarily long distances, even when the classical potential becomes negligible, and hence, their interaction can be described as nonlocal.

Such a non-locality is a phenomenon that is rather rare in physics. One can even raise serious objection to non-locality since, at first glance, it does not seem to be compatible with relativity due to the possibility of transmission of signals at faster than the speed of light. But contradiction with relativity does not seem to arise, because no useful signal can be transmitted this way. In any case, the existence of such non-locality has now been experimentally demonstrated beyond any reasonable doubt.

## 4. Quantum Entanglement

Customarily, a pair of photons is generated by parametric down conversion of a laser beam whereby the polarization of the original laser photon is mutually shared by the two resultant photons 1 and 2, making them maximally entangled since they share a conserved quantity. As a result of the interconnectedness brought about by the quantum potential, if the polarization of one of the particles is measured after separating them by an arbitrarily large distance, the other particle instantaneously reacts and possesses the complementary polarization necessary for conservation. We would then immediately know what measurement outcome will be obtained if we choose to measure particle 2.

Just prior to the measurement, however, the quantum potential of both particles is affected by the measuring system for example of particle 1. Therefore particle 2 being part of the overall quantum system is informed of the imminent act of the particular measurement of particle 1. After the measurement of particle 1, its $\psi$ field diffuses by interaction with the thermally active particles constituting the measuring device. But the $\psi$ field of particle 2 still persists along with its conserved property, which can then be measured to show the complementary conserved polarization. Alternatively, we can choose to use particle 2 for further maximal entanglement with a member of another pair of entangled particles. We can even store particle 2 in that state with the help of an optical delay line for entanglement with a member of a subsequently created pair of maximally entangled particles, as has been remarkably demonstrated by Megidish et al [5].

Roger Penrose states [6], "Since, according to quantum mechanics, entanglement is such a ubiquitous phenomenon—and we recall that the stupendous majority of quantum states are actually entangled ones—why is it something that we barely notice in our direct experience in the world?" In his opinion, Nature herself is continually enacting some state reduction process. We do find existence of relatively independent particles. When entanglement between the particles is lost, the wave function $\Psi$ for two particles can be factorized and written as a product

$$\Psi = \psi_1(X_1)\psi_2(X_2).$$

Then $$P = |\Psi|^2 = |\psi_1|^2 + |\psi_2|^2$$

And the quantum potential becomes a sum of two terms:

$$Q_1 = -\frac{\hbar^2}{2m} \frac{\nabla_1^2 R_1(X_1)}{R_1(X_1)},$$

$$Q_2 = -\frac{\hbar^2}{2m} \frac{\nabla_2^2 R_2(X_2)}{R_2(X_2)}.$$

Thus, each quantum potential is dependent only on the coordinates of a single particle causing each one to behave quasi independently.

## 5. A Heuristic Depiction

The mysterious non-locality of entanglement may be comprehensible in terms of the intrinsic quantum fluctuations of the underlying, indestructible, and immutable quantum fields that fill all space and time. As discussed earlier, the apparently chaotic quantum fluctuations in any space time element can be represented by a coherent wave function. Since the immutable magnitude of an underlying quantum field is the same throughout the universe, reflecting this reality the wave function $\psi_v$ representing its vacuum fluctuations in any space time element should be the same all over, resulting in an immense ensemble of identical quantum entities. Because of the plethora of interactions between the quantum fields predicted by QFT, there will be at least a minimal degree of mesoscopic entanglement [7, 8] between all the $\psi_v$ throughout the universe. To give just one example, the ubiquitous Higgs Ocean will interact with all the fields except those producing massless bosons. Of course, the number of interactions that contribute to various degrees of entanglement on a universal scale is beyond listing. Consequently, it should be possible to construct a *universal wave function* comprising at least the minimally entangled $\psi_v$ of all the space time elements.

When two particles are maximally entangled in a Bell state, their wave functions are not factorizable. But the wave function of each of the entangled particles, being also constituted from quantum fluctuations, can be superposed and entangled, by the overabundance of interactions between the quantum fields, with the *universal wave function* and thereby the complementary conserved properties of the two particles will remain in constant correlation even when they are separated in space by a great distance.

When a property of one particle is measured, the other particle instantaneously reacts because they are part of the *overall universal wave function that acts as a quantum mechanical Einstein-Rosen bridge [9] as envisioned by Maldacena and Susskind [10]*. As mentioned earlier, such an instantaneous interaction is consistent with relativity since it does not involve transmission of a useful signal

Still, non-locality of even random signals, signifying that space itself may be entangled seems rather aberrant to some eminent scientists, such as, Brian Green [10], John Bell [11], and S. Goldstein [13]. Therefore, a conjecture made by Bohm is worthy of further scrutiny. Bohm states [14] "Even in connection with gravitational theory, general relativity indicates that the limitation of speeds to the velocity of light does not necessarily hold universally. If we adopt the spirit of general relativity, which is to seek to make the properties of the matter that moves in this space, then it is quite conceivable that the metric, and therefore the limiting velocity, may depend on the $\psi$ field as well as on the gravitational tensor $g^{\mu\nu}$. In the classical limit the $\psi$-field could be neglected, and we would get the usual form of covariance. In any case, it can hardly be said that we have a solid experimental basis for requiring the same form of covariance at very short distances that we require at ordinary distances." Bohm's conjecture becomes significant in light of the temporal non-conservation of energy by all the quantum fluctuations at the core of the universe, which are also veiled in the sense that they cannot be observed without being disturbed.

In summary, since the underlying immutable quantum field fills all space with the same amount and the quantum fluctuations are correlated in a universal scale, the miracle of non-locality of quantum entanglement can possibly be comprehended as a natural process. In addition, the wave function and its mysterious collapse may also be understood in terms of the reality of the demonstrated fundamental structure of our universe, which is supported by quantum field theory.

## 6. Acknowledgement

The author wishes to thank Professor Zvi Bern for helpful discussions.